\documentclass{article}
\usepackage{geometry} % see geometry.pdf on how to lay out the page. There's lots.
\usepackage{amssymb}
\usepackage{amsmath}
\usepackage{authblk}
\usepackage{bbm}
\usepackage{graphicx}
\usepackage[T1]{fontenc}
\usepackage[utf8]{inputenc}
\usepackage{authblk}
\usepackage{epstopdf}
\usepackage{amsthm}

\def\independenT#1#2{\mathrel{\rlap{$#1#2$}\mkern2mu{#1#2}}}

\newtheorem{theorem}{Theorem}

\usepackage{authblk}
\allowdisplaybreaks

\begin{document}
\title{A Formal Causal Interpretation of the Case-Crossover Design}
\author[1]{Zach Shahn\thanks{zach.shahn@ibm.com}}
\author[2,3]{Miguel A. Hern\'an}
\author[2]{James M. Robins}
\affil[1]{IBM Research, Yorktown Heights, NY}
\affil[2]{Departments of Epidemiology and Biostatistics, Harvard University T H Chan School of Public Health, Boston, MA}
\affil[3]{CAUSALab, Harvard T.H. Chan School of Public Health, Boston, MA, U.S.A.}
\maketitle
%  This will produce the submission and review information that appears
%  right after the reference section.  Of course, it will be unknown when
%  you submit your paper, so you can either leave this out or put in 
%  sample dates (these will have no effect on the fate of your paper in the
%  review process!)

%  These options will count the number of pages and provide volume
%  and date information in the upper left hand corner of the top of the 
%  first page as in published papers.  The \pagerange command will only
%  work if you place the command \label{firstpage} near the beginning
%  of the document and \label{lastpage} at the end of the document, as we
%  have done in this template.

%  Again, putting a volume number and date is for your own amusement and
%  has no bearing on what actually happens to your paper!  

%  The \doi command is where the DOI for your paper would be placed should it
%  be published.  Again, if you make one up and stick it here, it means 
%  nothing!

%  This label and the label ``lastpage'' are used by the \pagerange
%  command above to give the page range for the article.  You may have 
%  to process the document twice to get this to match up with what you 
%  expect.  When using the referee option, this will not count the pages
%  with tables and figures.  

%  put the summary for your paper here

\begin{abstract}
The case-crossover design (Maclure, 1991) is widely used in epidemiology and
other fields to study causal effects of transient treatments on acute
outcomes. However, its validity and causal interpretation have only been
justified under informal conditions. Here, we place the design in a formal
counterfactual framework for the first time. Doing so helps to clarify its
assumptions and interpretation. In particular, when the treatment effect is
non-null, we identify a previously unnoticed bias arising from strong common
causes of the outcome at different person-times. We analyze this bias and demonstrate its potential importance with
simulations. We also use our derivation of the limit of the case-crossover
estimator to analyze its sensitivity to treatment effect heterogeneity, a
violation of one of the informal criteria for validity. The upshot of this
work for practitioners is that, while the case-crossover design can be
useful for testing the causal null hypothesis in the presence of baseline
confounders, extra caution is warranted when using the case-crossover design
for point estimation of causal effects.
\end{abstract}
%  Please place your key words in alphabetical order, separated
%  by semicolons, with the first letter of the first word capitalized,
%  and a period at the end of the list.
%

%  As usual, the \maketitle command creates the title and author/affiliations
%  display 

%  If you are using the referee option, a new page, numbered page 1, will
%  start after the summary and keywords.  The page numbers thus count the
%  number of pages of your manuscript in the preferred submission style.
%  Remember, ``Normally, regular papers exceeding 25 pages and Reader Reaction 
%  papers exceeding 12 pages in (the preferred style) will be returned to 
%  the authors without review. The page limit includes acknowledgements, 
%  references, and appendices, but not tables and figures. The page count does 
%  not include the title page and abstract. A maximum of six (6) tables or 
%  figures combined is often required.''

%  You may now place the substance of your manuscript here.  Please use
%  the \section, \subsection, etc commands as described in the user guide.
%  Please use \label and \ref commands to cross-reference sections, equations,
%  tables, figures, etc.
%
%  Please DO NOT attempt to reformat the style of equation numbering!
%  For that matter, please do not attempt to redefine anything!

\section{Introduction}
\label{s:intro}

The case-crossover design (Maclure, 1991) is used in epidemiology and other
fields to study causal effects of transient treatments on acute outcomes.
One of its major advantages is that it only requires
information from individuals who experience the outcome of interest (the
cases). Another appealing feature is that under certain circumstances (which
we will discuss at length) the case-crossover estimator adjusts for
unobserved time invariant confounding. In a seminal application of this
design (Mittleman et al., 1993), researchers obtained data on the physical
activity (a transient treatment) of individuals who experienced a myocardial
infarction (MI, an acute outcome). They then defined any person-times less
than one hour after vigorous activity as `treated', and all other
person-times as `untreated'. Finally, they considered each person-time as an
individual observation and computed a Mantel-Haenszel estimate of the
corresponding hazard ratio (Tarone, 1981; Nurminen, 1981; Kleinbaum et al.,
1982; Greenland and Robins, 1985). This hazard ratio estimate was
interpreted as the causal effect of vigorous physical activity on MI. Some
variants of the case-crossover design allow flexible control time selection
strategies where control times can follow outcome occurrence (e.g. Levy et
al., 2001), but in this paper we restrict attention to studies in which follow-up is terminated at the time of the first
outcome occurrence as in the above MI example.

Past authors have extensively considered several threats to validity of the
case-crossover design (Maclure, 1991; Greenland, 1996; Vines and Farrington,
2001; Levy et al., 2001; Janes et al., 2005; Mittleman and Mostofsky, 2014),
and conditions for causal interpretation of the estimator have been
informally stated in the literature. The usual criteria cited are that: (a)
the outcome has acute onset; (b) the treatment is transient; (c) there are
no unobserved post-baseline common causes of treatment and outcome; (d)
there are no time trends in treatment; and (e) the treatment effect is
constant across subjects.

%Condition (e) is a consequence of using the Mantel-Haenszel estimator of a common stratum-specific odds ratio. 
The Mantel-Haenszel estimator was originally applied to estimate the
treatment-outcome odds ratio when subjects were classified in strata sharing
values of confounders $V$, and observed subjects in each stratum could be
conceived of as independent draws from the (hypothetically) infinite stratum
population. Under the assumptions that stratum-specific odds ratios are all
equal and observations are independent within each stratum, the
Mantel-Haenszel estimator was proven consistent for the constant odds ratio
as the number of strata approach infinity even if only a few subjects are
observed in each stratum (Breslow, 1981). Since the values of the
confounders $V$ are held constant within each stratum, the constant odds
ratio can be endowed with a causal interpretation if $V$ includes all
confounders. The same goes for the rate ratio (Robins and Greenland, 1985).

Maclure's idea was to regard person-times (rather than subjects) as the
units of analysis and subjects as the strata, then apply the Mantel-Haenszel
estimator. As Maclure (1991) put it: ``In the case-crossover design, the
population base is considered to be stratified in the extreme, so there is
only one individual per stratum... Use of subjects as their own controls
eliminates confounding by subject characteristics that remain constant".
Analogy to past applications of the Mantel-Haenszel estimator would
seem to imply that the case-crossover design eliminates baseline confounding
as a source of bias assuming a constant treatment effect across subjects
(informal condition (e)) and independent identically distributed
observations across time within each subject. Of course, these two
assumptions are unlikely to be satisfied in most research settings: the
effect of treatment is rarely the same in all subjects, and variables at
different person-times are typically not independent within subjects.
Informal assumptions (a)-(d) can be viewed as a more plausible alternative
to independent person-times, but to determine when the
case-crossover estimator is asymptotically unbiased for causal effects in
the presence of unobserved confounding requires a formal analysis.

Here we place the case-crossover design in a formal counterfactual causal
inference framework (Rubin, 1978; Robins, 1986). Doing so helps to clarify
its assumptions and interpretation. In Section 2, we introduce notation,
describe the (possibly hypothetical) cohort that gives rise to the data in a
case-crossover analysis, and summarize the MI study in more detail so that
it can serve as a running example. In Section 3, we define a natural
estimand motivated by a hypothetical randomized trial practitioners of the
case-crossover design might wish to emulate. In Section 4, we state formal
assumptions (mostly analogous to informal assumptions (a)-(e)) that allow us
to causally interpret the limit of the case-crossover estimator and under
which the limit approximates the trial estimand from Section 3. We identify
and characterize a previously unnoticed bias present when there exist strong
common causes of the outcomes at different times (as would seem likely in
many instances) and the treatment effect is non-null. In Section 5, we
discuss this bias and illustrate it with simulations. We also use our
results from Section 4 to analyze sensitivity to effect heterogeneity, i.e.
violations of informal assumption (e). In Section 6, we conclude. Our
general message to practitioners is that, while the case-crossover can be a
clever way to test the null hypothesis of no causal effect in the presence
of unobserved baseline confounding, its point estimates of non-null effects
can be sensitive to violations of unrealistic assumptions.

\section{Data Generating Process}

\subsection{Notation}\label{cohort} 
While case-crossover studies only use data from subjects who
experience the outcome, we will nonetheless describe a full cohort from
which these subjects are drawn in order to facilitate the definition of
certain concepts and quantities of interest. Consider a cohort of
individuals followed from baseline (i.e. study entry)--defined by calendar
time, age, or time of some pre-defined index event--until they develop the
outcome or the administrative end of follow-up, whichever occurs first. For
simplicity, we assume no individual is lost to follow-up. Subjects are
indexed by $i$, $i\in \left\{ 1,...,N\right\} $. Subject $i$ is followed for
at most $T+1$ person-times (e.g. hours) indexed by $j\in \{0,\ldots ,T\}$.
For simplicity we take $T$ to be the same for all subjects. Let $A_{ij}$ be
a binary variable taking values 0 and 1 indicating whether subject $i$ was
treated at time $j$. Let $Y_{ij}$ be a binary variable taking values 0 and 1
indicating whether the outcome of interest occurred in subject $i$ before
time $j+1$. We assume that $Y_{ij}$ is a `time to event' outcome in the
sense that if $Y_{ij}=1$ then $Y_{ij^{\prime }}=1$ for all $j^{\prime }>j.$
The above implies the temporal ordering $A_{ij},Y_{ij},A_{i\left(
j+1\right)}.$ Thus the outcome has an acute onset as required by informal
condition (a). %If an outcome is recurrent, we can consider the time of its
%first occurrence. 
We define $A_{ij}=0$ if the event has occurred by time $j$.%
%, i.e. if $Y_{i\left( j-1\right) }=1.\ $

For a time-varying variable $X$, we denote by $\bar{X}_{ij}$ the history $%
(X_{i0},\ldots ,X_{ij})$ of $X$ in subject $i$ up (i.e. prior) to time $j+1$%
. We will often omit the subscript $i$ in the subsequent notation because we
assume the data from different subjects $i$ are independent and identically
distributed. Let $V$ denote a possibly multidimensional and unobserved
baseline confounding variable that we assume has some population density $%
p(v)$. (For notational convenience we shall write conditional probabilities $%
P\{\cdot |\cdot ,V=v$\} given $V=v$ as $p_{v}\{\cdot |\cdot $\}. To avoid
measure theoretic subtleties we shall henceforth assume that when $V$ has
continuous components, conditions sufficient to pick out a particular
version of $P\{\cdot |\cdot ,V=v$\} have been imposed as in Gill and Robins
(2001).) Let $\bar{U}_{T}$ denote common causes of outcomes (but not
treatments) at different person-times not included in $V$. For example, in
the MI and exercise study, $U_{j}$ could denote formation of a blood clot by
hour $j$ after baseline. We assume that the $N$ subjects are $iid$
realizations of the random vector $(V,\bar{A}_{T},\bar{Y}_{T},\bar{U}_{T})$
and that $U_j$ precedes $A_j$ and $Y_j$ in the temporal ordering at each $j$%
. Recall that in a case-crossover study the observed data on subject $i$ are 
$\left( \bar{A}_{iT},\bar{Y}_{iT}\right)$ as data on $V$ and $\bar{U}_{T}$
are not available.

We assume that the causal directed acyclic graph (DAG) (Greenland, Pearl,
and Robins, 1999) in Figure 1 describes the data generating process within
levels of baseline confounders $V$. This DAG encodes aspects of informal
assumptions (b) and (c). One salient feature of the DAG is that there are no
directed paths from a current treatment to an outcome at a later time that
do not first pass through the outcome at the time of the current treatment
or through a later treatment. This can be considered a representation of
informal assumption (b) that treatment is transient. The DAG also excludes
any common causes of treatments and outcomes other than $V$ not through past
outcomes. (Since occurrence of the outcome at time $j$ determines the values
of all variables at all later time points, outcome variable nodes in the DAG
trivially must have arrows to all temporally subsequent variables.) This
represents informal assumption (c) which bars non-baseline confounding. This
DAG also has fully forward connected treatments with arbitrary common causes
of treatment at different times $\bar{U}_{AT}$, indicating that we put no $%
causal$ restrictions on the treatment assignment process. (We will, however,
impose distributional assumptions.) We provide a fuller discussion of causal
assumptions in Section 4, but we find it helpful to keep this DAG in mind.

\begin{figure}[h]
\centering
\includegraphics[scale=.2]{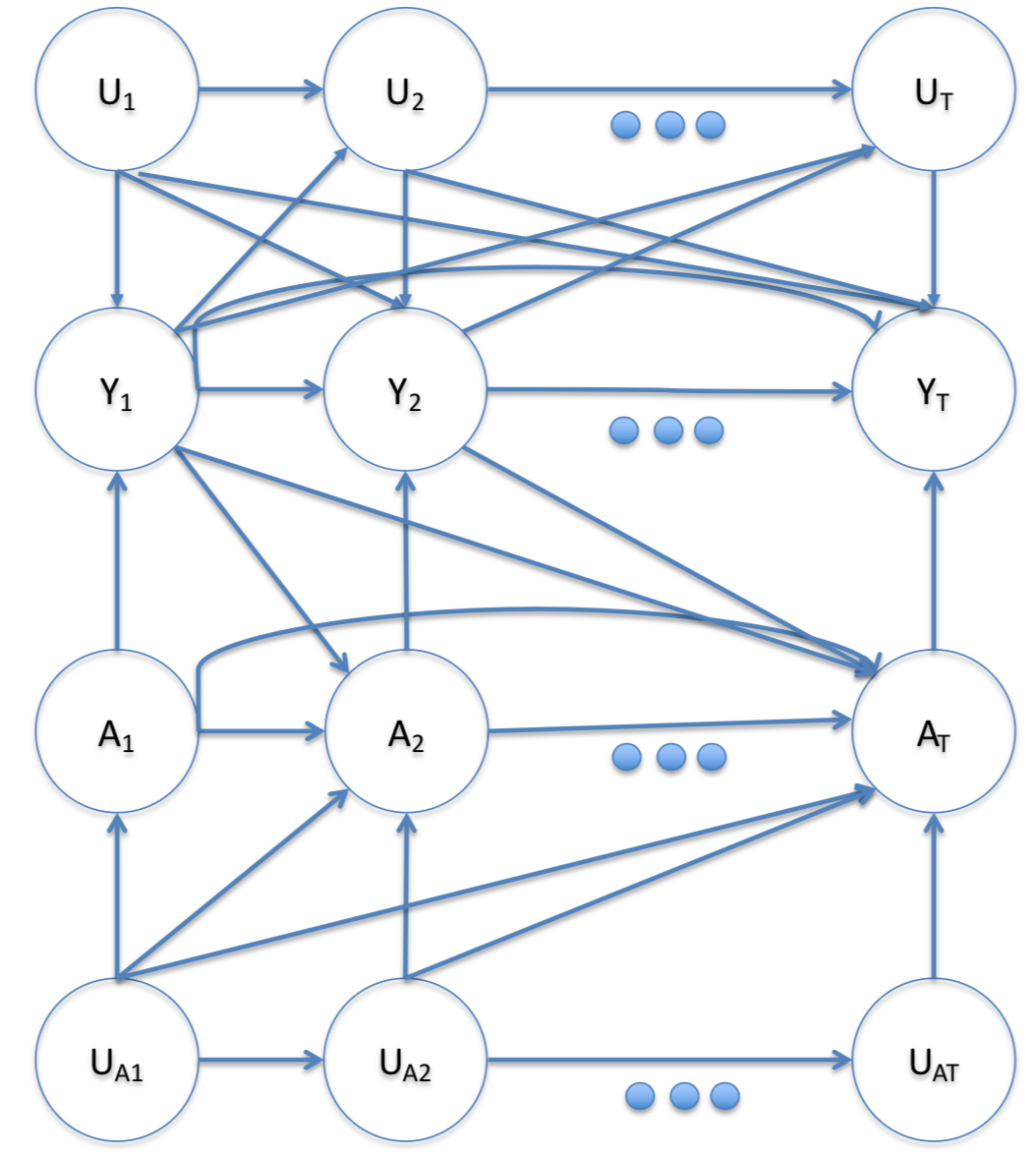}
\caption{Causal DAG within levels of $V$. A $V$ node with arrows pointing
into every other node was omitted for visual clarity. This figure appears in
color in the electronic version of this article, and any mention of color
refers to that version.}
\label{fig2}
\end{figure}

\subsection{The Case-Crossover Design}
The outcome-censored case-crossover Mantel-Haenszel estimator requires data from subjects who experience the
outcome on treatment status at the time of outcome occurrence and at
designated `control' times preceding the outcome. It is computed as
follows:

\begin{itemize}
\item Select a random sample of $H$ person-times from the $H^{\ast}$ person
times $ij$ satisfying $Y_{ij}=1$, $Y_{i\left( j-1\right) }=0$, and $j>W$
where $W$ is a maximum `look back' time chosen by the investigator. We refer
to these $H^{\ast}$ person-times when the outcome occurred for the first
time and after time $W$ as the set of `case' person-times.
\item Let $i_{h}j_{h}$ denote the person-time of the $h^{th}$ element of the
set of $H$ sampled case person times. From the same subject $i_{h}$, select $%
m$ times $\{j_{h}-c_1,\ldots ,j_{h}-c_{m}\}$ from the $W$ times prior to the
time $j_{h}$ of subject $i_{h}^{\prime }s$ first outcome event. We call
these $m$ times the `control' person-times for subject $i_{h}$. We discuss
selection of `control' times below.
\item Let $A_{h}^{1}$ denote the treatment at the case time and $%
(A_{h1}^{0},\ldots ,A_{hm}^{0})$ denote treatments at the $m$ control times
in subject $i_{h}$. The Mantel-Haenszel case-crossover estimator $\widehat{%
IRR}_{MH}$ is
\begin{equation}  \label{estimator}
\widehat{IRR}_{MH}=\frac{\sum_{h}\sum_{l=1}^{m}\mathbbm{1}%
\{A_{h}^{1}=1,A_{hl}^{0}=0\}}{\sum_{h}\sum_{l=1}^{m}\mathbbm{1}%
\{A_{h}^{1}=0,A_{hl}^{0}=1\}}.
\end{equation}%
Note that for subject $i_{h}$ the only data necessary to compute $\widehat{%
IRR}_{MH}$ is $(A_{h}^{1},A_{h1}^{0},\ldots ,A_{hm}^{0}).$
\end{itemize}

Intuitively, the more subjects tend to be treated at the time of the outcome
but not at earlier control times as opposed to vice versa, the stronger the estimated effect
of treatment. To fix ideas, we consider an example of a case-crossover study from the
literature. In a simplified version of Mittleman et al.'s (1993) study on
the impact of exercise on MI mentioned in the introduction, suppose we
collect data from a random sample of patients suffering MI on a particular
Sunday. We record whether each patient exercised in the hour immediately
preceding their MI and whether they exercised in the same hour the day
before their MI. We compute the Mantel-Haenszel
case-crossover estimator (\ref{estimator}): in the numerator is the number
of subjects who exercised immediately prior to their MI but not 24 hours
before, and in the denominator is the number of subjects who did not
exercise immediately prior to their MI but did 24 hours before. Mittleman et
al. estimated a ratio of 5.9 (95$\%$ CI 4.6,7.7). They found the ratio was
much higher among subjects who rarely exercised (107, 95$\%$ CI 67,171) than
those who exercised regularly (2.4, 95$\%$ CI 1.5,3.7) prior to the study
period.

Many approaches to selecting control times might be acceptable. In the MI example, the lookback window
is the 24 hours before the MI and there is only one control time exactly 24
hours before the outcome time. So $W=24$, $m=1$,
and $c_1=24$ in the notation above.

\section{A Natural Estimand}\label{trial} 

Consider $T$ parallel group randomized trials in which, in
trial $k$, treatment is randomly assigned at and only at time $k$ to all
subjects who have yet to experience the outcome. Such a time $k$-specific
trial could estimate the immediate effect of treatment at time $k$. 
%This (discrete-time) hazard ratio or relative risk, assumed constant over $k=1,\ldots,T$, could be a natural quantity for the case-crossover estimator to target.   
To formalize, we adopt the counterfactual framework of Robins (1986). Let $%
Y_{ij}^{\bar{a}_{j}}$ be the value of the outcome at time $j$ had, possibly
contrary to fact, subject $i$ followed treatment regime $\bar{a}_{j}\equiv
(a_{1},\ldots ,a_{j})$ through time $j$. We refer to $Y_{ij}^{\bar{a}_{j}}$
as a counterfactual or potential outcome. Since we will frequently consider
treatment interventions at a single time point, we also introduce the
notation $Y^{a_j}$ as shorthand for $Y^{\bar{A}_{j-1},a_j}$, i.e. the
counterfactual value of random variable $Y_j$ under observed treatment
history through $j-1$ and treatment at time $j$ set to $a_j$. The randomized
trial described above conducted at time $k$ would yield an estimate of
relative risk or hazard ratio $\rho_k\equiv P(Y_k^1=1|\bar{Y}%
_{k-1}=0)/P(Y_k^0=1|\bar{Y}_{k-1}=0)$. In
the remainder of the paper, we will establish (strong) assumptions under
which the case-crossover estimator approximately converges to a causally
interpretable quantity. Under these assumptions, $\rho_k$ does not depend
on $k$ and the case-crossover estimator approximately approaches
\begin{equation}  \label{natural_estimand}
\rho \equiv \rho_k \text{ constant over }k.
\end{equation}
%We consider the impact of violations of some of these assumptions.  

\section{Derivation of the Counterfactual Interpretation of the Limit of the
Case-Crossover Estimator}\label{derivation}

\subsection{Assumptions}\label{assumptions} 
Our goal is to specify natural and near minimal
assumptions that allow us to causally interpret the limit of the
case-crossover estimator. 
%Most of our assumptions can be interpreted as formalizations of informal assumptions (a)-(e). 
Counterfactuals and the observed data are linked by the following standard
assumption: 
\begin{equation}  \label{consistency}
\textbf{Consistency: } Y_{j}^{\bar{A}_{j}}=Y_{j}\text{ for all }j.
\end{equation}
Consistency states that the counterfactual outcomes corresponding to the
observed treatment regimes are equal to the observed outcomes. Consistency
is a technical assumption that has no counterpart in the informal
assumptions (a)-(e) but is implicit in almost all analyses.

We assume that the causal graph in Figure 1 describes the data generating
process (Greenland, Pearl, and Robins, 1999). We will state some specific
assumptions implied by the graph in counterfactual notation and also state
additional assumptions. Figure 1 encodes informal assumption (c) that there
are no post-baseline confounders not contained in $V$, i.e. 
\begin{equation}  \label{seq_ex}
\textbf{Sequential Exchangeability: }A_{j}\mathpalette{\protect%
\independenT}{\perp}\left\{ Y_{k}^{\bar{a}_{k}};k\geq j\right\} |\bar{Y}%
_{j-1}=0,\bar{A}_{j-1}=\bar{a}_{j-1},V=v\text{ for all }j.
\end{equation}
%Note that sequential exchangeability implies $U_j\mathpalette{\protect\independenT}{\perp}A_j|\bar{Y}_{j-1},\bar{A}_{j-1},V$, a fact we make use of later (and is apparent in Figure 1). 
See Appendix 2 for further details. An example violation of (\ref{seq_ex})
in the MI study would be if caffeine intake at hour $j$ both encouraged
exercise and increased MI risk at $j$. We might expect that confounders of
this sort in the MI study (short term encouragements to exercise that are
associated with MI) are weak.

The DAG in Figure 1 also reflects informal assumption (b) that effects are
transient by implying that $A_j$ has no direct effect on $Y_{j+1},\ldots,Y_T$
not through $A_{j+1},\ldots,A_T$ and $Y_j$ for all $j$. Graphically, this is
the statement that the only treatment variable that is a parent of $Y_j$ is $%
A_j$. We might hope that the graphical definition of the transient effect
assumption would be equivalent to the assumption that, conditional on $V$,
counterfactual hazards are independent of past treatment history, i.e. that $%
\lambda^{\bar{a}_j}_{vj}\equiv p_v(Y_j^{\bar{a}_j}=1|\bar{Y}_{j-1}=0)$ does
not depend on $\bar{a}_{j-1}$. However, this is not generally true due to
collider bias (Hernan et al., 2004) stemming from selection on survival and
the presence of common causes $V$ and $\bar{U}_{T}$ of the outcome in Figure
1 since, e.g., the path $A_{j-1} \rightarrow Y_{j-1} \leftarrow U_{j-1}
\rightarrow Y_j$ is open. Because of the collider bias, a formal
counterfactual definition of the transient effect assumption requires that
we condition on $U$-histories. Specifically, let $\lambda_{vj}^{a_j}(\bar{a}%
_{j-1},\bar{u}_j)$ denote $p_v(Y_j^{a_j}=1|\bar{Y}_{j-1}=0,\bar{A}_{j-1}=%
\bar{a}_{j-1},\bar{U}_j=\bar{u}_j)$, i.e. the conditional counterfactual
hazard at time $j$ under treatment $a_j$ given past treatments $\bar{a}_{j-1}
$, common causes of outcomes $\bar{u}_j$, and baseline confounders $v$. As
in Figure 1, we assume: 
\begin{equation}  \label{ptni}
\textbf{UV-Transient Hazards: } \lambda_{vj}^{a_j}(\bar{A}_{j-1}=\bar{a}%
_{j-1},\bar{U}_{j-1}=\bar{u}_j) \text{ does not depend on } \bar{a}_{j-1}.
\end{equation}%
That is, conditional on $V$ and the history of $U$, the current
counterfactual hazard does not depend on past treatments. This assumption is
consistent with the absence of any mention of such dependence in the case
crossover literature. Biological considerations determine the plausibility
of (\ref{ptni}). In the MI study, (\ref{ptni}) would be violated if exercise can have delayed
effects on MI. Maclure (1991) argued that delayed effects
would be weak in this setting.

Under (\ref{ptni}), we can write the counterfactual hazard $\lambda
_{vj}^{a_{j}}(\bar{a}_{j-1},\bar{u}_j)$ for any $\bar{a}_{j-1}$ as $\lambda
_{vj}^{a_{j}}(\bar{u}_j)$. The causal hazard ratio at time $j$ given $\bar{u}%
_j$ and $v$ is then $\lambda _{vj}^{1}(\bar{u}_j)/\lambda_{vj}^{0}(\bar{u}_j)
$. We assume that the causal hazard ratio is constant: 
\begin{align}  \label{estimand}
\begin{split}
&\textbf{Constant Causal Hazard Ratio: } \\
& \beta \equiv \beta_{vj}(\bar{u}_j)\equiv \lambda _{vj}^{1}(\bar{u}%
_j)/\lambda _{vj}^{0}(\bar{u}_j)\text{ does not depend on }v, j, \text{ or } 
\bar{u}_j.
\end{split}%
\end{align}
(\ref{estimand}) is a version of the constant effects assumption (e). Under
the constant hazard ratio assumption (\ref{estimand}), $\beta=\rho$ from (%
\ref{natural_estimand}) and the trial described in Section \ref{trial} would
target $\beta$. 
%(\ref{estimand}) makes stronger homogeneity assumptions than (\ref{natural_estimand}), which only assumes constant hazard ratios at each time as opposed to constant hazard ratios conditional on time and the unmeasured variables $v$ and $\bar{u}_j$.  
(Note that for (\ref{estimand}) not to depend on the specific set of
variables included in $V$ and $\bar{U}$, which we leave unspecified,
requires that $\beta_{vj}(\bar{u}_j)$ is collapsible over $\bar{U}$ and $V$.
While it is well known that hazard ratios are not generally collapsible
(Greenland, 1999), the scenario in which non-collapsibility arises entails a
baseline exposure influencing failure at all future timepoints. Under our
transient effects assumption, $\beta_{vj}(\bar{u}_j)$ is just an immediate
conditional relative risk, which is collapsible.) (\ref{estimand}) is a very
strong assumption unlikely to ever hold exactly. Violations can be less
extreme in subpopulations, e.g. subjects who exercise regularly in the MI
study. We examine sensitivity to violations of (\ref{estimand}) in Section %
\ref{heterogeneity}.

Under the above assumptions, we will show (in Theorem 1) that the
case-crossover estimator approaches $\beta$ times a multiplicative bias
term. The assumptions below, invoked in the order they are introduced, are
sufficient to ensure the multiplicative bias term is near 1. 
\begin{equation}  \label{rare_outcome}
\textbf{Rare Outcome: } \prod_{j=1}^T(1-\lambda_{vj}^{a_j}(\bar{u}%
_j))<\epsilon\text{ }\forall \bar{u}_T,v,\bar{a}_T \text{ and }\epsilon 
\text{ a small positive number.}\ 
\end{equation}
This rare outcome assumption holds under all levels of $V$, $\bar{U}$, and $%
\bar{A}$. Because $(V,\bar{U})$ can be very high dimensional and contain
post-baseline information, it is unlikely that this assumption holds in the
MI study. For example, formation of a clot might cause a violation. But we
will see that bias can be small even if this assumption fails as long as
cases occurring under the violating $(V,\bar{U})$ levels do not account for
a large proportion of total cases.

Next, we require the assumption 
\begin{equation}\label{time_mod}
\begin{split}
& \textbf{No Time-Modified Confounding:} \\
& \sum_{l=1}^m\sum_{k>W}\int_{v}\lambda
_{vk}^{0}\{p_{v}(A_{k}=1,A_{k-c_l}=0)-p_{v}(A_{k}=0,A_{k-c_l}=1)\}p(v)dv \\
& =\sum_{l=1}^m\sum_{k>W}\Big[\int_{v}\lambda _{vk}^{0}p(v)dv\times
\int_{v}p_{v}(A_{k}=1,A_{k-c_l}=0)-p_{v}(A_{k}=0,A_{k-c_l}=1)p(v)dv\Big]
\end{split}%
\end{equation}%
where $\lambda _{vk}^{0}$ is the untreated counterfactual hazard at $k$
marginal over $\bar{u}_{k}$ and $k-c_l$ is a control time for an outcome occurring
at $k$. A sufficient condition for (\ref{time_mod}) to hold is that, for
each $k$ and $l$ the marginal correlation $Cov({p_{V}(A_{k}=1,A_{k-c_l}=0)-p_{V}(A_{k}=0,A_{k-c_l}=1)},\lambda_{Vk}^{0})$ is zero between the random functions $p_{V}(A_{k}=1,A_{k-c_l}=0)-p_{V}(A_{k}=0,A_{k-c_l}=1)$ and $\lambda _{Vk}^{0}$
of $V$. In fact, we require only that the sum over $k$ and $l$ of the $k$-specific
covariances for each control time is zero. This condition prevents bias
from so called \textit{time modified} baseline confounders $V$ (Platt et
al., 2009) which, by definition, are baseline confounders $V$ that predict
both (i) the hazard of an unexposed subject failing at various times $k$
and (ii) the difference in marginal probabilities of the events $(A_{k}=0,A_{k-c_l}=1)$ and $(A_{k}=1,A_{k-c_l}=0)$. The case-crossover literature distinguishes between baseline and post-baseline confounders and
says the former are allowed but not the latter. The more relevant
distinction is whether a confounder has time-varying effects. To understand
the issue, first consider a post-baseline confounder. We gave the example
earlier of caffeine intake ($C_{k}$) at time $k$ impacting probability of
both exercise and MI at $k$ (more precisely, between $k$ and $k+1$). $C_{k}$
is temporally a post-baseline variable as its value is realized at time $k$,
but in the causal ordering it could be equivalent to a baseline variable if
it is not influenced by past treatments. For example, coffee at time $k$
could be equivalent to a $k$-hour delayed release caffeine pill at baseline.
Suppose $Z(k)\in V$ is a baseline variable (like the delayed release
caffeine pill) such that $Z(k)=1$ causes $A_{k}=1$ and $Y_{k}=1$ to be more
likely. $Z(k)$ would induce bias just like $C_{k}$, even though $Z(k)\in V$
is a baseline confounder that (unlike $C_{k}$) would not lead to a violation
of (\ref{seq_ex}). However, whenever $Z(k)=1$, $p(A_{k}=1,A_{k-c_l}=0)-p(A_{k}=0,A_{k-c_l}=1)$ and $\lambda _{vk}^{0}$ will both be large, inducing a correlation of the sort banned by (\ref{time_mod}).
Thus, (\ref{time_mod}) serves to ban time modified confounding. 

There is a straightforward intuitive motivation behind informal assumption
(d) that there are no time trends in treatment. Because control times always
precede case times, a steady change in treatment probability over time would
result in a preponderance of discordant pairs of one type over the other in
the estimator (\ref{estimator}) even in the absence of any causal effect of
treatment on outcome. Our version of informal assumption (d) is: 
\begin{equation}\label{pairwise}
\begin{split}
& \textbf{No Time Trends in Treatment:}\sum_{l=1}^m\sum_{k>W}p(A_{k}=1,A_{k-c_l}=0)=%
\sum_{l=1}^m\sum_{k>W}p(A_{k}=0,A_{k-c_l}=1)\\
&\text{for all } k,c_l\text{ such that }k-c_l\text{ would be a control time if the outcome were
to occur at }k.
\end{split}%
\end{equation}%
Note that we make this assumption marginally over $V$ and $\bar{U}_{k}$. A sufficient condition for the No Time Trends assumption to hold is that, at every time $k$ and for every control time $k-c_l$, the assumption $p(A_{k}=1,A_{k-c_l}=0)=p(A_{k}=0,A_{k-c_l}=1)$ of (marginal) pairwise exchangeability holds as previously derived by Vines and Farrington (2001).
%Note further that under our Rare Outcome assumption the above assumption is (essentially) equivalent to the assumption $Pr(A_{k}=1,A_{k-c}=0|\bar{Y}_{k-1}=0)=Pr( A_{k}=0,A_{k-c_l}=1|\bar{Y}_{k-1}=0)$. 
(\ref{pairwise}) can be viewed as a more precise formulation of the informal \textquotedblleft no time trends in treatment\textquotedblright assumption (d). Exposure
can exhibit arbitrarily complex temporal dependence (as illustrated in the
DAG in Figure 1) as long as (\ref{pairwise}) holds. Whether this
assumption holds depends in part on how control times are chosen. In the MI
study, control times 12 hours prior to the outcome could be
much less likely to satisfy (\ref{pairwise}) than control times 24 hours
prior (e.g. 2PM the previous day would be a better control time than 2AM the
morning of an MI that occurred at 2PM). 
%The vaccine study would violate (\ref{pairwise}) if vaccination probability is associated with age.
%\begin{remark}
%It may appear odd that many of our assumptions are framed in terms of variables $V$ and $\bar{U}$ that are not precisely defined. $V$ can be any sufficient adjustment set and $\bar{U}$ the corresponding remaining common causes of the outcome. Thus, $V$ might be as large as the baseline state of the universe or as small as a minimal adjustment set. 
%\end{remark}
%Note that (\ref{no_trends}) is assumed to hold within each level of $V$ but still allows for other baseline variables that are not confounders to possibly lead to differences between the probabilities of $(A_k=0,A_{c_k}=1)$ and $(A_k=1,A_{c_k}=0)$. For example, in the MI study, having made a dentist appointment prior to baseline for time $k$ but not time $c_k$ would make exercise at $c_k$ but not $k$ much more likely than vice versa. But (assuming that dentist appointments do not also cause MI and are therefore not included in $V$) this would not violate (\ref{no_trends}).  While a stronger assumption than strictly necessary (see Remark 2), (\ref{no_trends}) does still allow for high levels of pre-baseline determination of treatment.
%A strained and purely illustrative example, perhaps the lifestyle of a stereotypical American football fan (beer, red meat) is a confounder that also makes exercise during the afternoon on Sunday (when most football games are played) less likely than Saturday in the MI study. 

We note that all the results in the paper hold if there exist any $V$ and $\bar{U}$ for which
both the independencies of the causal DAG in Figure 1 and assumptions (\ref{consistency})-(\ref{pairwise}) are satisfied.  
%We utilize the above assumptions to derive a causal interpretation of the limit of the Mantel-Haenszel estimator.
%In the vaccine
%example, all days between 365 and 42 days prior to the outcome on which the
%subject is at least six months old are controls. Since $m_h$ depends on subject age under this scheme, a slight modification of our analysis is necessary to accommodate the vaccine example.

\subsection{The limit of the Mantel-Haenszel estimator}

\label{derivation} In the theorem below we derive the limit of $\widehat{IRR}%
_{MH}$ (\ref{estimator}) in the outcome-censored case-crossover design under
an asymptotic sequence in which the full cohort, the number of cases in the
cohort, and the number of sampled cases grow at similar rates, i.e. $%
N\rightarrow \infty ,H^{\ast }/N\rightarrow d_{1}>0,$ and $H/H^{\ast
}\rightarrow d_{2}>0$. We also assume subjects are $iid$.
The proof is in Appendix 1. 
\begin{theorem}
(i) Assume consistency (\ref{consistency}), exchangeability (\ref{seq_ex}), UV-transient hazards (\ref{ptni}), and constant hazard ratio $\beta$ (\ref{estimand}). Then, under the outcome-censored case-crossover design, $\widehat{IRR}_{MH} \overset{p}{\rightarrow } \beta\tau$ with multiplicative bias term
\begin{equation*}
\tau = \frac{\sum_{l=1}^m\int_v \sum_{k>W}\sum_{\bar{u}_k}\alpha_{vkl}(\bar{u}_k,1,0)p(v)dv}{\sum_{l=1}^m\int_v \sum_{k>W}\sum_{\bar{u}_k}\alpha_{vkl}(\bar{u}_k,0,1)p(v)dv}
\end{equation*}
where $\alpha_{vkl}(\bar{u}_k,a,a')= \lambda_{vk}^0(\bar{u}_k)p_v(\bar{Y}_{k-1}=0,A_k=a,A_{k-c_l}=a',\bar{U}_k=\bar{u}_k)$.\\
(ii) Under additional assumptions Rare Outcome (\ref{rare_outcome}), No Time-Modified Confounding (\ref{time_mod}), and No Time Trends in Treatment (\ref{pairwise}), $\tau \approx 1$.
\end{theorem}
\section{Analysis of selected sources of bias}
\subsection{Bias due to strong common causes of the outcome}\label{common_cause_bias} 
As discussed earlier, our rare outcome assumption
within levels of (possibly post-baseline and high dimensional) common causes
of the outcome is novel and unreasonably strong. In this subsection, we will
examine analytically and through simulations the bias that arises when it
fails. We first consider the special case in which, at each time $k$, exposure is determined by an independent coin flip with success probability $p$. In
that case, as shown in Remark 1 in Appendix 1, the multiplicative bias $\tau$ from Theorem 1 is well
approximated by 
\begin{equation}\label{simple_bias_analysis}
\frac{\int_{v}\sum_{k>W}%
\sum\limits_{\bar{u}_{k}}M_{v}(\bar{u}_{k})\left\{ 1-\lambda
_{v,k-c}^{0}(\bar{u}_{k-c})\right\} p(v)dv}{\int_{v}\sum_{k>W}\sum%
\limits_{\bar{u}_{k}}M_{v}(\bar{u}%
_{k})\left\{ 1-\lambda _{v,k-c}^{1}(\bar{u}_{k-c})\right\} p(v)dv}
\end{equation}
where $M_{v}(\bar{u}_{k})=\lambda _{vk}^{0}(\bar{u}_{k})\left\{
\prod_{j=1}^{k}p_{v}(u_{j}|\bar{Y}_{j-1}=0,\bar{U}_{j-1}=\bar{u}%
_{j-1})\right\} $. %Note that each term in the numerator and denominator is
%non-negative.%\begin{align*}
%&G_v(a,a',\bar{a}_{k/k,k-c},\bar{u}_k)\equiv p_v(A_k=a|\bar{Y}_{k-1}=0,A_{k-c}=a^{'},\bar{A}_{k/k,k-c}=\bar{a}_{k/k,k-c},\bar{U}_k=\bar{u}_k)\times\\
%&\prod_{s=k-c+1}^{k-1}p_v(a_s|\bar{Y}_{s-1}=0,A_{k-c}=a^{'},\bar{A}_{s-1/k-c}=\bar{a}_{s-1/k-c},\bar{U}_{k-c}=\bar{u}_{k-c})\times\\
%&p_v(A_{k-c}=a^{'}|\bar{Y}_{k-c-1}=0,\bar{A}_{k-c-1}=\bar{a}_{k-c-1},\bar{U}_{k-c-1}=\bar{u}_{k-c-1})\prod_{s=1}^{k-c-1}p_v(a_s|\bar{Y}_{s-1}=0,\bar{A}_{s-1}=\bar{a}_{s-1},\bar{U}_s=\bar{u}_s).
%\end{align*}

Disparities between the numerator and denominator of the bias term (\ref%
{simple_bias_analysis}) will lead to bias of the estimator. Before
examining disparities related to non-negligible $V$ and $U$-specific survival
probabilities, we note that the bias contribution of a disparity at a given
level of $v$ and $\bar{u}_{k}$ depends on the weight $M_{v}(\bar{u}_{k})p(v)$,
which is large when both the probability of observing $(v,\bar{u}_{k})$
and the probability of an untreated event occurring at $k$ given $v$ and $%
\bar{u}_{k}$ are large. Thus, the larger the proportion of total cases
occurring at $v$ and $\bar{u}_{k}$, the more that failure of the rare
outcome assumption at $v$ and $\bar{u}_{k}$ biases the estimator.

The only difference between the numerator and denominator of (\ref%
{simple_bias_analysis}) is that where $1-\lambda _{v,k-c}^{0}(\bar{u}_{k-c})
$ appears in the numerator, $1-\lambda _{v,k-c}^{1}(\bar{u}_{k-c})$ appears
in the denominator. The ratio of the term in the numerator to that in
the demoninator is $\frac{1-\lambda _{v,k-c}^{0}(\bar{u}_{k-c})}{1-\beta \lambda _{v,k-c}^{0}(\bar{u}%
_{k-c})}$. When $\beta =1,$ this
factor is equal to 1 and there is no bias. When $\beta \neq 1$, the bias
is away from the null since $\frac{1-\lambda _{v,k-c}^{0}(\bar{u}_{k-c})}{1-\beta \lambda _{v,k-c}^{0}(\bar{u}%
_{k-c})}>1$ if and only if $\beta >1$
and thus the MH estimator converges to a limit that is further from 1 than
the true $\beta $ and in the same direction. For the mutiplicative bias to
be nonnegligible requires a violation of the rare outcome assumption in
which there exist histories $v$, $\bar{u}_{k}^{\ast }$ for which both $M_{v}(\bar{%
u}_{k}^{\ast })/\sum\limits_{\bar{u}_{k}}M_{v}(\bar{u}_{k})$ and $\beta
\lambda _{v,k-c}^{0}(\bar{u}_{k-c}^{\ast })$ are non-negligible. 

We illustrate this bias with a simulation. For $N=100,000$ subjects, we
simulated treatments and counterfactual outcomes for 24 time steps or until the first occurrence
of the outcome according to the following data generating process (DGP). 
\begin{equation*}
\begin{split}
& U_{t}\sim Bernoulli(.001);\text{ }\lambda
_{t}^{0}(U_{t-1},U_{t})=min(1/2,.45U_{t-1}+.45U_{t}) \\
& Y_{t}^{0}\sim Bernoulli(\lambda _{t}^{0}(U_{t-1},U_{t}));\text{ }\lambda
_{t}^{1}(U_{t-1},U_{t})=2\lambda _{t}^{0}(U_{t-1},U_{t}) \\
& Y_{t}^{1}\sim Bernoulli(\lambda _{t}^{1}(U_{t-1},U_{t}));\text{ }A_{t}\sim
Bernoulli(.5);\text{ }Y_{t}=A_{t}Y_{t}^{1}+(1-A_{t})Y_{t}^{0}
\end{split}%
\end{equation*}%
\begin{figure}[h]
\centering
\includegraphics[scale=.15]{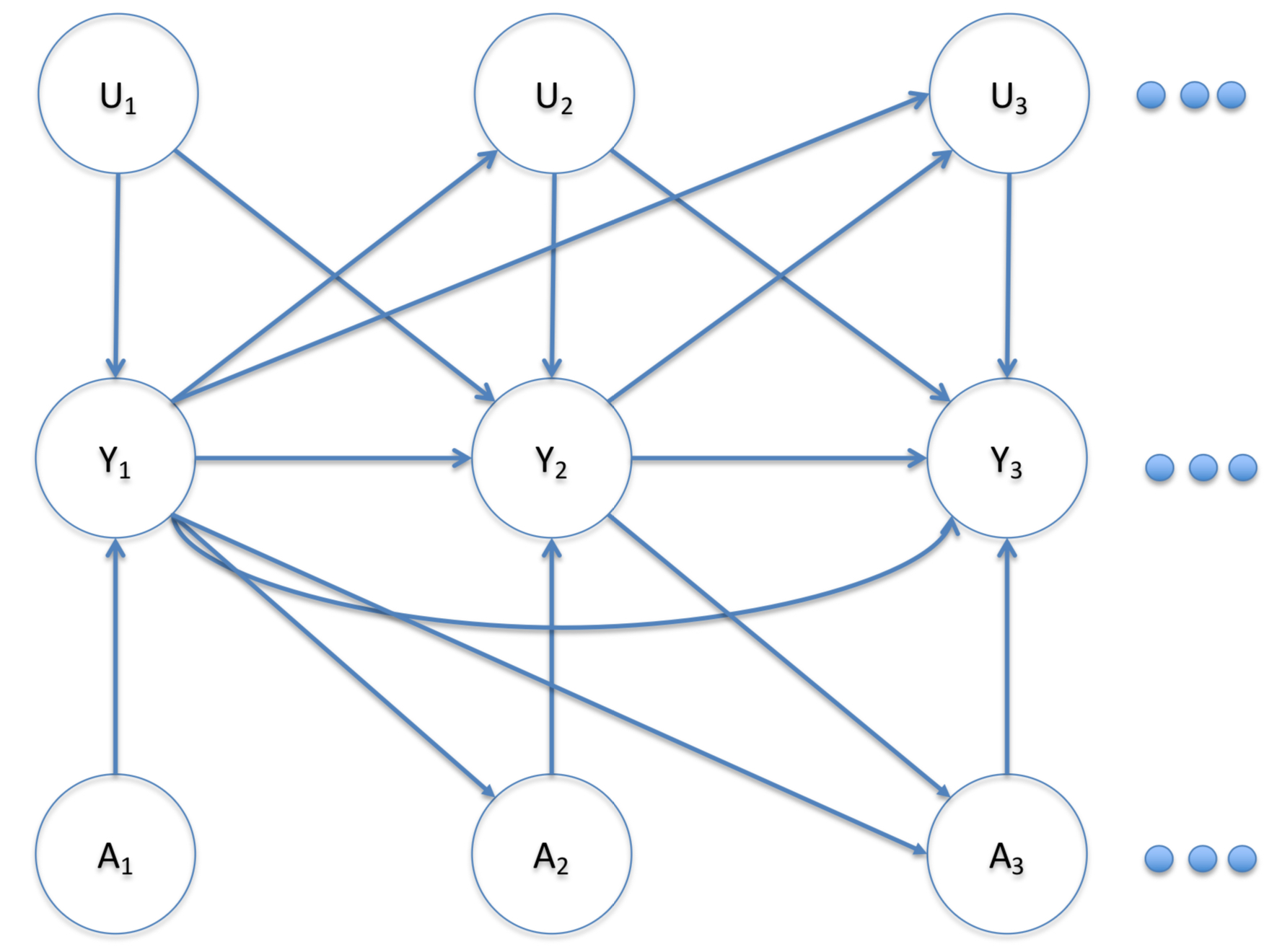}  
\caption{Causal DAG for simulation DGP with unobserved post-baseline common
causes of outcomes at different times}
\label{fig3}
\end{figure}
The DAG for this DGP is depicted in Figure 2. The true value of $\beta $ is
2. There are no common causes of treatments and outcomes, treatments are
independent identically distributed and hence exhibit no time trends, and
the outcome is rare when marginalized over $U$. (While the outcome is not rare
when $U_{t}=1$, it is rare that $U_{t}=1$.) Yet the limit of the
case-crossover estimator using the time prior to outcome occurrence as the
control is approximately 2.8. The estimator fails because the outcome was
common when $U_t$ or $U_{t-1}$ were 1 and a
large proportion of total cases occurred when $U_t$ or $U_{t-1}$ were 1. The bias is away from
the null, as predicted by our analysis above. The effect of $U$ on the
outcome needed to be strong to produce the bias in this simulation. If $%
\lambda _{t}^{0}(U_{t-1},U_{t})=min(1/2,.25U_{t-1}+.25U_{t})$ instead of $%
min(1/2,.45U_{t-1}+.45U_{t})$, then the case-crossover estimator is
about 2.3 instead of 2.8. A recently formed blood clot could roughly
play the role of $\bar{U}$ in the MI example--a rare event that does not influence
probability of exposure, greatly increases probability of the outcome at
multiple time points after the clot forms, and without which the outcome is
rare. 

Now we consider bias in the more general scenario where treatments are correlated across time. In Appendix 1 we expand the bias term $\tau$ as 
\begin{equation}\label{bias_analysis_form}
\frac{\int_v \sum_{k>W}\sum\limits_{\bar{u}_k}M_v(\bar{u}_k)(1-\lambda_{v,k-c}^{0}(\bar{u}_{k-c}))\sum\limits_{\bar{a}_{k/k,k-c}}G_v(1,0,\bar{a}_{k/k,k-c},\bar{u}_k)\prod\limits_{s\neq k-c,k}(1-\lambda_{vs}^{a_s}(\bar{u}_s))p(v)dv}{\int_v \sum_{k>W}\sum\limits_{\bar{u}_k}M_v(\bar{u}_k)(1-\lambda_{v,k-c}^{1}(\bar{u}_{k-c}))\sum\limits_{\bar{a}_{k/k,k-c}}G_v(0,1,\bar{a}_{k/k,k-c},\bar{u}_k)\prod\limits_{s\neq k-c,k}(1-\lambda_{vs}^{a_s}(\bar{u}_s))p(v)dv}
\end{equation}
where $\bar{a}_{k/k,k-c}$ denotes $\bar{a}_k$ excluding $a_k$ and $a_{k-c}$ and $G_v(a,a',\bar{a}_{k/k,k-c},\bar{u}_k)$ (defined in Appendix 1) roughly corresponds to the probability of observing treatment trajectory with $a_k=a$, $a_{k-c}=a'$, and treatment at the other time points equal to $\bar{a}_{k/k,k-c}$. When treatments are correlated, $G_{v}(1,0,\bar{a}_{k/k,k-c},\bar{u}_{k})$ in the numerator might assign high weights to different treatment sequences $\bar{a}_{k/k,k-c}$ than $G_{v}(0,1,\bar{a}_{k/k,k-c},\bar{u}_{k})$ in the denominator, and under failure of the rare outcome assumption the highly weighted treatment sequences in the numerator might have
significantly different survival probabilities ($\prod\limits_{s\neq
k-c,k}(1-\lambda _{vs}^{a_{s}}(\bar{u}_{s}))$) for some values of $v$ and $\bar{u}_k$ than the highly weighted
treatment trajectories in the denominator. By the reasoning we applied to infer direction of bias in the case with uncorrelated exposures, strongly weighted untreated survival probabilities in the numerator combined with strongly weighted treated survival probabilities in the denominator would lead to bias away from the null, and vice versa. Depending on the treatment correlation pattern, treated or untreated survival probabilities might be more strongly weighted in the numerator or denominator.  Thus, in the correlated treatment case the resulting bias can be either toward or away from the null. As in the case without correlated exposures, the magnitude of the bias contribution stemming from this dynamic for a given $v$ and $\bar{u}_k$ depends on $M_v(\bar{u}_k)$. 

To illustrate, we modify our previous simulation example to add correlations in treatments across time. If time bins are interpreted as hours in the previous simulation, they are seconds in this one. Exposure and the unobserved common cause of the outcome are still independently assigned to one hour intervals as in the previous simulation. This induces perfect correlation between treatments corresponding to one second time bins within the same hour. The untreated one second discrete hazards are set to preserve the hourly untreated survival probability from the previous simulation, and the multiplicative treatment effect within each one second bin is again set to 2. To formalize, we simulated data according to
\begin{equation*}
\begin{split}
& \tilde{U}_{k}\sim Bernoulli(.001)\text{ for }k\in\{1,\ldots,24\}; U_{kt}=\tilde{U}_{k}\text{ for }k\in\{1,\ldots,24\}, t\in\{1,\ldots,3600\}\\
& \tilde{A}_{k}\sim Bernoulli(.5)\text{ for }k\in\{1,\ldots,24\}; A_{kt}=\tilde{A}_{k}\text{ for }k\in\{1,\ldots,24\}, t\in\{1,\ldots,3600\}\\
& \lambda_{kt}^{0}(\bar{U}_{kt})= 0.000166(U_{kt} + U_{k-1t}); Y_{kt}^{0}\sim Bernoulli(\lambda_{kt}^{0}(\bar{U}_{kt}));\text{ }\lambda_{kt}^{1}(\bar{U}_{kt})=2\lambda_{kt}^{0}(\bar{U}_{kt}) \\
& Y_{kt}^{1}\sim Bernoulli(\lambda_{kt}^{1}(\bar{U}_{kt}));\text{ }Y_{kt}=A_{kt}Y_{kt}^{1}+(1-A_{kt})Y_{kt}^{0}
\end{split}%
\end{equation*}%
where we have indexed `hours' by $k$ and seconds within hours by $t$. The true value of $\beta$ in this DGP is again 2, but the
case-crossover estimate using the time bin exactly one hour (3600 seconds)
prior to the case as the control (as in the previous simulation) is 1.84. So
modifying the DGP from the previous simulation to be finer grained (and thus inducing correlations between treatments across times) while still
constructing the case-crossover estimator in the identical way made the bias
switch direction. While neither DGP (fine or coarse) is
likely to be a $good$ approximation to any realistic process, we would argue
that it is difficult to reason about which would be a $better$ approximation
for any given use case with the broad characteristics that both simulations
share. Thus, the two simulations taken together illustrate that bias from
strong common causes of the outcome, when present, can be both sizable and
unpredictable. (See Web Appendix A for analytic confirmation of simulation results from both DGPs using (\ref{mult_bias_term}), discussion of what drives the discrepancy between the two simulations, and further analysis of bias in the correlated exposure setting.)
%We can directly understand how $U$ impacts the case-crossover estimator (\ref{estimator}) by reducing the frequency of discordant pairs in which the control time is exposed, thus increasing the estimator by decreasing the count in its denominator. The sequence $(U_{t-1}=1,A_{t-1}=1,Y_{t-1}=0,A_{t}=0,Y_{t}=1)$ is particularly unlikely, since $Y_{t-1}$ is likely to equal 1 when $U_{t-1}=1$ and $A_{t-1}=1$ as a result of $U_{t-1}=1$ increasing the baseline risk of $Y_{t-1}$. And because $U_{t-1}=1$ also increases the probability of $Y_{t}=1$, there will be many cases contributing to the estimator in which this dynamic was at play.  

%{\LARGE Zach Say something about the papers saying case crossover biased
%with nonsymetrical data ie those analyses are conditonal on treatment rther
%than marginalizing over it (Maybe put this in discussion)}

\subsection{Treatment Effect Heterogeneity}\label{heterogeneity} 
We now examine sensitivity to violations of the
constant causal hazard ratio assumption if the rare outcome assumption
holds. For simplicity, we consider a scenario where there are just two types
of subjects and counterfactual hazard ratios are constant across time within
types. For $g \in \{0,1\}$, say subjects of type $g$ arise from the
following data generating process: 
\begin{align*}
\begin{split}
A_1,\ldots,A_T \overset{iid}{\sim} Bernoulli(p_{A,g}); \text{ }%
Y^0_1,\ldots,Y^0_T\overset{iid}{\sim} Bernoulli(\lambda^0_{g}) \\
Y^1_1,\ldots,Y^1_T\overset{iid}{\sim} Bernoulli(\lambda^1_{g}); \text{ }Y_j
= A_jY^1_j + (1-A_j)Y^0_j
\end{split}%
\end{align*}
with data censored at the first occurrence of the outcome. So within each
type $g$, the constant causal hazard ratio is $\lambda^1_{g}/\lambda^0_{g}$.
Let $p_g$ denote the proportion of the population of type $g=1$ at baseline,
which under the rare outcome assumption would also be approximately the
proportion of type $g=1$ among surviving subjects at all subsequent followup
times. According to equation (\ref{step3}) from the proof of Theorem 1, if
the rare outcome assumption holds then the case-crossover estimator with m=1
(i.e. using just one control) will approach 
\begin{equation}  \label{example_limit}
\frac{\lambda^1_{g=1}p_{A,g=1}(1-p_{A,g=1})p_g +
\lambda^1_{g=0}p_{A,g=0}(1-p_{A,g=0})(1-p_g)}{%
\lambda^0_{g=1}p_{A,g=1}(1-p_{A,g=1})p_g +
\lambda^0_{g=0}p_{A,g=0}(1-p_{A,g=0})(1-p_g)}.
\end{equation}
(\ref{example_limit}) can be expressed as a weighted average of $%
\lambda^1_{g=0}/\lambda^0_{g=0}$ and $\lambda^1_{g=1}/\lambda^0_{g=1}$, 
\begin{equation*}
\frac{\lambda^1_{g=0}}{\lambda^0_{g=0}} \frac{\delta}{\delta+\theta} + \frac{%
\lambda^1_{g=1}}{\lambda^0_{g=1}} \frac{\theta}{\delta+\theta},
\end{equation*}
where $\delta = \lambda^0_{g=0}p_{A_g=0}(1-p_{A,g=0})(1-p_g)$ and $\theta =
\lambda^0_{g=1}p_{A,g=1}(1-p_{A,g=1})p_g$. Hence, the limit of the
case-crossover estimator is bounded by the group-specific hazard ratios.

The relative risk computed from any of the RCTs described in Section \ref%
{trial} would approach 
\begin{equation}  \label{estimand2}
\frac{\lambda^1_{g=1}p_g + \lambda^1_{g=0}(1-p_g)}{\lambda^0_{g=1}p_g +
\lambda^0_{g=0}(1-p_g)}.
\end{equation}
Like the case-crossover limit, the RCT estimand can be expressed as a
weighted average of $\lambda^1_{g=0}/\lambda^0_{g=0}$ and $%
\lambda^1_{g=1}/\lambda^0_{g=1}$: 
\begin{equation}
\frac{\lambda^1_{g=0}}{\lambda^0_{g=0}}\frac{\lambda^0_{g=0}(1-p_g)}{%
\lambda^0_{g=0}(1-p_g) + \lambda^0_{g=1}p_g} + \frac{\lambda^1_{g=1}}{%
\lambda^0_{g=1}}\frac{\lambda^0_{g=1}p_g}{\lambda^0_{g=0}(1-p_g) +
\lambda^0_{g=1}p_g}.
\end{equation}
Without loss of generality assume $\frac{\lambda^1_{g=0}}{\lambda^0_{g=0}}>%
\frac{\lambda^1_{g=1}}{\lambda^0_{g=1}}$. The ratio of the weight placed on
the higher hazard ratio to the weight placed on the lower hazard ratio in
the RCT estimand is 
\begin{equation}  \label{rct}
\gamma_{RCT}\equiv\frac{\lambda^0_{g=0}(1-p_g)}{\lambda^0_{g=1}p_g}.
\end{equation}
The corresponding case-crossover weight ratio is 
\begin{equation}  \label{hetero_bias}
\gamma_{CC}\equiv\frac{\lambda^0_{g=0}(1-p_g)p_{A,g=1}(1-p_{A,g=1})}{%
\lambda^0_{g=1}p_gp_{A,g=0}(1-p_{A,g=0})}=\gamma_{RCT}\times\frac{%
p_{A,g=0}(1-p_{A,g=0})}{p_{A,g=1}(1-p_{A,g=1})}.
\end{equation}
(\ref{hetero_bias}) implies that bias of the case-crossover estimator due to
treatment effect heterogeneity depends on the difference in treatment
probability between groups with different effect sizes. If treatment
probability does not vary across groups with different treatment effects,
effect heterogeneity will not induce bias in the case-crossover estimator.
When treatment probabilities do vary, whichever group has higher treatment
variance $p_{A,g}(1-p_{A,g})$, i.e. whichever group has probability of
treatment closer to .5, will be weighted too highly by the case-crossover
estimator compared to the RCT estimand. Some intuition behind this behavior
is that the closer the treatment probability within a group is to .5, the
more subjects from that group will contribute discordant case-control pairs
to the case-crossover estimator, weighting the estimator disproportionately
toward the effect within that group.

For illustrative purposes, consider a numerical example where we set: 
\begin{align*}
\lambda^0_{g=0}=.001; \text{ }\lambda^1_{g=0}=.002; \text{ }%
\lambda^0_{g=1}=.0005; \text{ }\lambda^1_{g=1}=.005; \text{ }p_{A,g=0} = .8; 
\text{ }p_{A,g=1} = .5; \text{ }p_g=.5.
\end{align*}
Then $\lambda^1_{g=1}/\lambda^0_{g=1}=10$, $\lambda^1_{g=0}/\lambda^0_{g=0}=2
$, and the estimand (\ref{estimand2}) is equal to $4.67$. $\widehat{IRR}_{MH}
$ converges to 5.5, while the naive cohort hazard ratio estimator $\frac{%
P(Y=1|A=1)}{P(Y=1|A=0)}$ that does not adjust for the confounder $g$
approaches 4.9. In this example, bias from effect heterogeneity overrides
any benefits from control of unobserved confounding. (While we set baseline
outcome risks to be different across levels of $g$ in this example, note
that (\ref{hetero_bias}) implies this plays no role in inducing bias due to
effect heterogeneity.)

\begin{figure}[h]
\centering
\includegraphics[scale=.2]{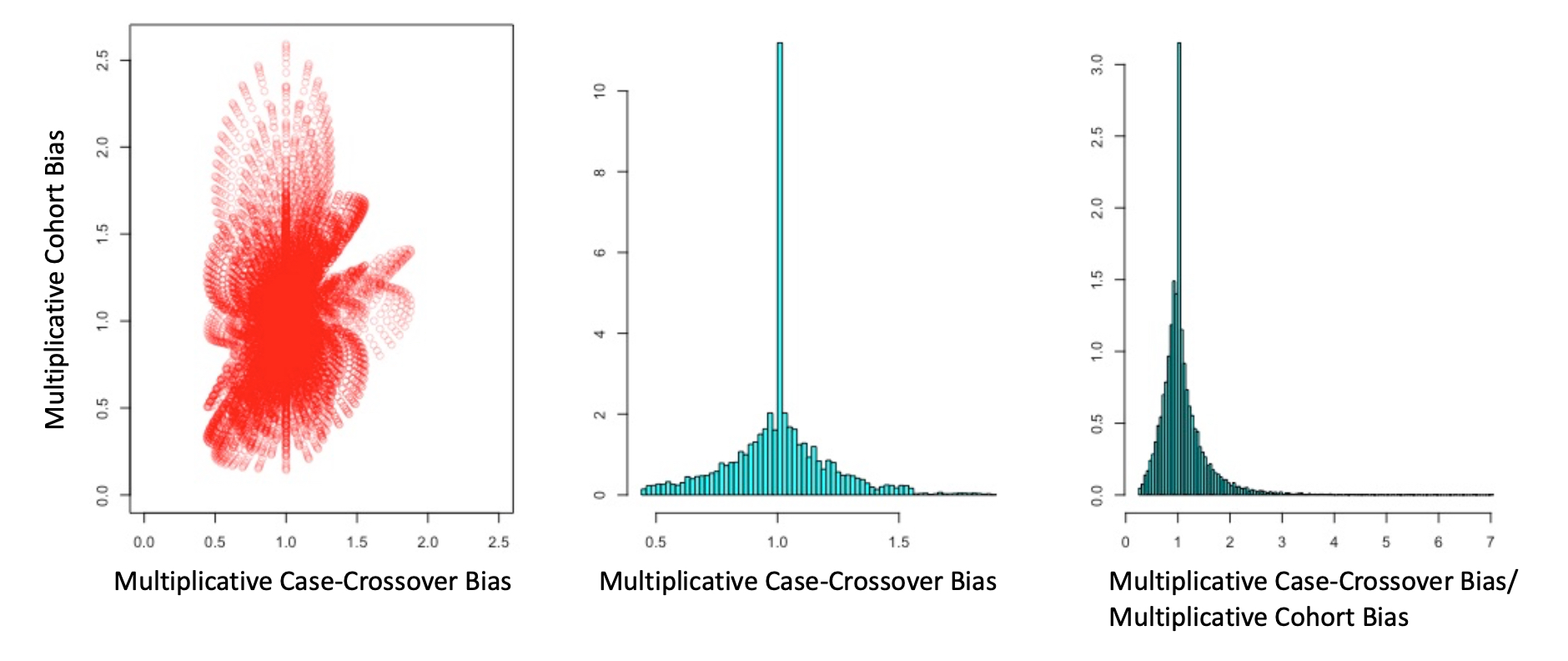}
\caption{Left: Scatterplot of case-crossover vs cohort estimator
multiplicative bias across a range of settings. Middle: Distribution of
case-crossover estimator bias across settings. Right: Distribution of ratio
of case-crossover bias to cohort bias across settings.}
\label{fig4}
\end{figure}

The specific numerical example above is a cautionary tale illustrating the
potential significance of heterogeneity induced bias. But if both cohort and
case-crossover analyses are feasible with available data, and unobserved
baseline confounding and effect heterogeneity vary within realistic ranges,
does one estimator tend to be more biased than the other? We addressed this
question in the framework of our toy example by computing the limiting
values of case-crossover and cohort estimators for a large grid of data
generating process parameter settings. We let $\lambda^0_{g=0}$ and $%
\lambda^0_{g=1}$ take values in $\{0.0005, 0.001\}$, $\lambda^1_{g=0}/%
\lambda^0_{g=0}$ take values in $\{1,\ldots,5\}$, $\lambda^1_{g=1}/%
\lambda^0_{g=1}$ take values in $\{1\times\lambda^1_{g=0}/\lambda^0_{g=0},%
\ldots,10\times\lambda^1_{g=0}/\lambda^0_{g=0}\}$, and $p_{A,g=0}$ and $%
p_{A,g=1}$ take values in $\{1/20,\ldots,19/20\}$. Figure \ref{fig4} shows
that neither estimator has a general advantage over the other across
parameter settings.

%Treatment effect heterogeneity was present in the MI study. %In the vaccine example, only one strain of the vaccine seemed to significantly increase the risk of meningitis. 
In the MI study, the effect of exercise appeared much greater in subjects
who rarely exercised than in those who exercised regularly. Probability of
treatment (i.e. exercise) clearly varied considerably between regular and
rare exercise groups. 
%, and probability of exercise in a given hour would be less than .5 for almost all subjects in both groups. 
Hence, we would expect an estimate of the marginal effect to be biased. The
authors of the MI study reported separate effect estimates for the strata
over which the effect was thought to vary. This is appropriate, as marginal
effect estimates for the full population can be misleading.

We note that the numerical analyses in this section provide a framework for quantitative bias analysis (Lash et al, 2014) to assess sensitivity to violations of the effect homogeneity assumption. Given a case-crossover estimate, an analyst can first specify a simple heterogeneity model (or many) similar to our toy example above. The analyst can then derive from (\ref{step3}) the expression for the limit of the case-crossover estimator under that model as a function of its parameters, just as we easily derived (\ref{example_limit}). Finally, by exploring a grid of plausible model parameters, the analyst can identify a range of true effect sizes that might result in the observed case-crossover estimate under effect heterogeneity. 

\section{Discussion}
We have put the case-crossover estimator on more solid theoretical footing
by providing a proof of its approximate convergence to a formal
counterfactual causal estimand, $\beta$, under certain assumptions. This
result alone may not be of much utility, but it was overdue for such a
widely used method. And the derivation yielded some practical insights as
byproducts.

First, we discovered a new source of potential bias when the treatment
effect is not null--strong common causes of the outcome across time. We
analyzed this bias and illustrated its potential significance and
unpredictability with simulations. The effect of the common cause needs to
be quite strong to induce sizable bias, but the fact that $(V,\bar{U})$ can
be high dimensional and temporally post-baseline increases the likelihood of
this in a real analysis. Formation of a blood clot might induce a bias of
this sort in the MI example, but it is difficult to speculate about how
often meaningful bias of this type appears in practice.

Second, expression (\ref{step3}) characterizing the limit of the
case-crossover estimator allowed us to quantify sensitivity to violations of
the constant treatment effect assumption. We analyzed a simple scenario with
two groups of subjects having potentially different baseline risks, exposure
rates, and treatment effects. The limit of the case-crossover estimator was
a weighted average of the group-specific hazard ratios. The bias relative to
the estimand (\ref{natural_estimand}) that would be targeted by a RCT
depends on the exposure rates in the groups. If the groups have the same
exposure rate, effect heterogeneity would not induce any bias. Otherwise,
whichever group had exposure rate closer to 0.5 would be overweighted. We
provided a numerical example in which significant unobserved baseline
confounding (which could be controlled by the case-crossover estimator) and
effect heterogeneity were both present. In this example, the effect
heterogeneity bias in the case-crossover estimator was greater than the
confounding bias in a standard cohort hazard ratio estimator, illustrating
that effect heterogeneity can sometimes override benefits from control of
unobserved baseline confounding in the case-crossover estimator. More
extensive numerical analyses showed that neither the cohort estimator nor
the case-crossover had a general advantage across a range of settings in
which the levels of unobserved confounding and effect heterogeneity varied.
An analyst concerned about bias from effect heterogeneity could employ the
general framework of our numerical studies to conduct a quantitative bias
analysis (Lash et al, 2014). 

Overall, the formal assumptions required for consistency mostly mapped onto
informal assumptions (a)-(e). Unsurprisingly for a method that has been used
for thirty years, our contributions do not drastically alter its recommended
use. As an illustrative exercise, we assess our simplified version of Mittelman
et al.'s (1993) study of the effect of exercise on MI assumption by
assumption through the lens of our analysis in Web Appendix B. 

We might summarize our general guidance to practitioners and consumers of
case-crossover analyses as follows. If unobserved baseline confounding is
thought to be serious and/or data collection for a cohort study is
unfeasible, the case-crossover should be considered as an option. If
interest lies only in testing the null hypothesis of no effect, fewer
assumptions are necessary. Under the null: the transient treatment
assumption automatically holds; common causes of the outcome do not induce
bias; the rare outcome assumption is not necessary; and there is no
treatment effect heterogeneity. Hence, the case-crossover design remains a
clever method for causal null hypothesis testing in the presence of
unmeasured baseline confounders under the exchangeability (\ref{seq_ex}), no
time trends in treatment (\ref{pairwise}), and no time-modified confounding (%
\ref{time_mod}) assumptions. If interest lies in obtaining a point estimate,
results should be interpreted with considerable additional caution as effect
heterogeneity, delayed treatment effects, and common causes of outcomes will
all be present to some degree, and as we have shown can have a large impact
on results. 
%It is admittedly more difficult to reason directly about the formal assumptions we presented than the informal ones. However, the causal DAG in Figure 1 provides a useful and intuitive tool for practitioners to assess the plausibility of these assumptions. Whenever this DAG describes the data generating process, one can assume effects are transient and there is no confounding. If, in addition, the outcome is rare, the marginal probability of treatment is the same at all case and control person-times (i.e. there are not time trends in treatment), and there are no common causes of the outcome, then the case-crossover estimator would be consistent for a (null or non-null) constant causal hazard ratio. These conditions are not harder to verify with subject matter knowledge than the informal assumptions and have the advantage of being more precise. In Web Appendix D, we provide an illustrative assessment of the simplified MI study through the lens of the framework developed here.

There are many variants of the case-crossover design, of which we have here only analyzed arguably the simplest one. One important extension of the MH estimator adjusts for post-baseline confounders through matching. Another variant employs conditional logistic regression in place of the MH estimator. In this case, Vines and Farrington (2001) showed that $joint$ exchangeability is required among all control times and the case time as opposed to just pairwise exchangeability. Additionally, in situations where time trends in treatment are present, the case-time-control method (Suissa, 1995) is often utilized and requires alternative assumptions (Greenland, 1996). The case-crossover design is also frequently applied in air pollution epidemiology. In this setting, the treatment regime is shared among all subjects and later values of treatment are not influenced by past values of subjects' outcomes, allowing more flexible control time selection strategies, including using control times following outcome occurrence (Navidi, 1998; Levy et al., 2001; Janes et al., 2005). It would be interesting to investigate these variants in a similar counterfactual framework.
\section*{Acknowledgements}
This research was partly funded by NIH grant R37 AI102634. We also wish to thank Sonia Hernandez-Diaz and Anke Neumann for guidance and helpful discussions. \vspace*{-8pt}

\section*{Code}
Code for the simulations and numerical study from Section 5 is available at $https://github.com/zshahn/case_crossover/$.
%  If your paper refers to supplementary web material, then you MUST
%  include this section!!  See Instructions for Authors at the journal
%  website http://www.biometrics.tibs.org

\section*{Appendix 1: Proof of Theorem 1}
\small 
(i) From the definition of $\widehat{IRR}_{MH}$, it is clear that under the
case-crossover design 
\begin{equation}
\widehat{IRR}_{MH}\overset{p}{\rightarrow }\frac{%
\sum_{l=1}^{m}Pr_{MH}(A^{1}=1,A_{l}^{0}=0)}{%
\sum_{l=1}^{m}Pr_{MH}(A^{1}=0,A_{l}^{0}=1)},  \label{limit}
\end{equation}%
where $Pr_{MH}(A^{1}=a,A_{l}^{0}=a^{\prime })$ is the probability under the
case-crossover sampling scheme that a randomly selected subject who
experienced an occurrence of the outcome at least W+1 time steps after
baseline will have treatment level $a$ at the time of the outcome and the $%
l^{th}$ selected `control' time from the duration $W$ lookback period
preceding the event in the same subject will have treatment level $a^{\prime
}$. For simplicity, we present the proof for the case where $m=1$ and, if
the outcome occurs at time $k$, there is one control time $k-c$. $W=c$ in
this scenario. According to this sampling scheme and letting $\bar{a}%
_{k/k,k-c}$ denote $\bar{a}_{k}$ excluding $a_{k}$ and $a_{k-c}$, we can
express (\ref{limit}) in terms of population probabilities as 
\begin{align}
& \frac{Pr_{MH}(A^{1}=1,A^{0}=0)}{Pr_{MH}(A^{1}=0,A^{0}=1)}  \label{step1} \\
& =\frac{\int_{v}\sum_{k>W}\sum_{\bar{a}_{k/k,k-c},\bar{u}_{k}}p_{v}(Y_{k}=1,%
\bar{Y}_{k-1}=0,A_{k}=1,A_{k-c}=0,\bar{A}_{k/k,k-c}=\bar{a}_{k/k,k-c},\bar{U}%
_{k}=\bar{u}_{k})p(v)dv}{\int_{v}\sum_{k>W}\sum_{\bar{a}_{k/k,k-c},\bar{u}%
_{k}}p_{v}(Y_{k}=1,\bar{Y}_{k-1}=0,A_{k}=0,A_{k-c}=1,\bar{A}_{k/k,k-c}=\bar{a%
}_{k/k,k-c},\bar{U}_{k}=\bar{u}_{k})p(v)dv}  \label{step2} \\
& =\frac{\int_{v}\sum_{k>W}\sum_{\bar{a}_{k/k,k-c},\bar{u}_{k}}\beta _{vk}(%
\bar{u}_{k})\lambda _{vk}^{0}(\bar{u}_{k})p_{v}(\bar{Y}%
_{k-1}=0,A_{k}=1,A_{k-c}=0,\bar{A}_{k/k,k-c}=\bar{a}_{k/k,k-c},\bar{U}_{k}=%
\bar{u}_{k})p(v)dv}{\int_{v}\sum_{k>W}\sum_{\bar{a}_{k/k,k-c},\bar{u}%
_{k}}\lambda _{vk}^{0}(\bar{u}_{k})p_{v}(\bar{Y}_{k-1}=0,A_{k}=0,A_{k-c}=1,%
\bar{A}_{k/k,k-c}=\bar{a}_{k/k,k-c},\bar{U}_{k}=\bar{u}_{k})p(v)dv}
\label{step3} \\
& =\beta \frac{\int_{v}\sum_{k>W}\sum_{\bar{u}_{k}}\lambda _{vk}^{0}(\bar{u}%
_{k})\sum_{\bar{a}_{k/k,k-c}}p_{v}(\bar{Y}_{k-1}=0,A_{k}=1,A_{k-c}=0,\bar{A}%
_{k/k,k-c}=\bar{a}_{k/k,k-c},\bar{U}_{k}=\bar{u}_{k})p(v)dv}{%
\int_{v}\sum_{k>W}\sum_{\bar{u}_{k}}\lambda _{vk}^{0}(\bar{u}_{k})\sum_{\bar{%
a}_{k/k,k-c}}p_{v}(\bar{Y}_{k-1}=0,A_{k}=0,A_{k-c}=1,\bar{A}_{k/k,k-c}=\bar{a%
}_{k/k,k-c},\bar{U}_{k}=\bar{u}_{k})p(v)dv}  \label{step4} \\
& =\beta \frac{\sum_{k}\int_{v}\sum_{\bar{u}_{k}}\lambda _{vk}^{0}(\bar{u}%
_{k})p_{v}(\bar{U}_{k}=\bar{u}_{k},A_{k-c}=0,A_{k}=1,\bar{Y}_{k-1}=0)p(v)dv}{%
\sum_{k}\int_{v}\sum_{\bar{u}_{k}}\lambda _{vk}^{0}(\bar{u}_{k})p_{v}(\bar{U}%
_{k}=\bar{u}_{k},A_{k-c}=1,A_{k}=0,\bar{Y}_{k-1}=0)p(v)dv}.
\label{mult_bias_term}
\end{align}%
\noindent We go from (\ref{step1}) to (\ref{step2}) by basic probability
rules; (\ref{step2}) to (\ref{step3}) by consistency (\ref{consistency}),
Sequential Exchangeability (\ref{seq_ex}), and UV-transient hazards (\ref%
{ptni}); (\ref{step3}) to (\ref{step4}) by Constant Hazard Ratio (\ref%
{estimand}); and (\ref{step4}) to (\ref{mult_bias_term}) by the law of total probability. 

{\small (ii) Under rare outcome assumption (\ref{rare_outcome}), $p_{v}(\bar{%
U}_{k}=\bar{u}_{k},A_{k-c}=a,A_{k}=a^{\prime },\bar{Y}_{k-1}=0)\approx p_{v}(%
\bar{U}_{k}=\bar{u}_{k},A_{k-c}=a,A_{k}=a^{\prime })$. And by (\ref{rare_outcome}) and the DAG in
Figure 1, $p_{v}(\bar{U}_{k}\approx\bar{u}_{k},A_{k-c}=a,A_{k}=a^{\prime })=p_{v}(\bar{U}_{k}=\bar{u}_{k})p_{v}(A_{k-c}=a,A_{k}=a^{\prime })$. Therefore, we
can approximate the bias term in (\ref{mult_bias_term}) as 
\begin{equation}\label{rare_bias}
\frac{\int_{v}\sum_{k>W}\sum_{\bar{u}_{k}}\lambda _{vk}^{0}(\bar{u}%
_{k})p_{v}(\bar{U}_{k}=\bar{u}_{k})p_{v}(A_{k}=1,A_{k-c}=0)p(v)dv}{%
\int_{v}\sum_{k>W}\sum_{\bar{u}_{k}}\lambda _{vk}^{0}(\bar{u}_{k})p_{v}(\bar{%
U}_{k}=\bar{u}_{k})p_{v}(A_{k}=0,A_{k-c}=1)p(v)dv}\approx\frac{\int_{v}\sum_{k>W}%
\lambda _{vk}^{0}p_{v}(A_{k}=1,A_{k-c}=0)p(v)dv}{\int_{v}\sum_{k>W}\lambda
_{vk}^{0}p_{v}(A_{k}=0,A_{k-c}=1)p(v)dv}. 
\end{equation}
Now, by applying assumption (\ref{time_mod}) and then (\ref{pairwise}), 
\begin{equation}
\int_{v}\sum_{k>W}\lambda
_{vk}^{0}\{p_{v}(A_{k}=1,A_{k-c}=0)-p_{v}(A_{k}=0,A_{k-c}=1)\}p(v)dv\approx
0,
\end{equation}%
where we again used the rare outcome assumption to approximate $\lambda^0_{vk}(\bar{u}_k)$ by the density $f^0_{vk}(\bar{u}_k)$. This implies that (\ref{rare_bias}) is approximately 1, proving the result. 
%where $\alpha_{vk}(\bar{u}_k)$ denotes $\lambda_{vk}^0(\bar{u}_k)\sum_{\bar{a}_{k/k,k-c}}p_v(\bar{Y}_{k-1}=0,A_k=1,A_{k-c}=0,\bar{A}_{k/k,k-c}=\bar{a}_{k/k,k-c},\bar{U}_k=\bar{u}_k)$.
}

\textbf{Remark}: In the absence of the rare disease assumption we can expand $\tau$ as
\begin{equation}\label{bias_analysis_formula}
\frac{\int_v \sum_{k>W}\sum\limits_{\bar{u}_k}M_v(\bar{u}_k)(1-\lambda_{v,k-c}^{0}(\bar{u}_{k-c}))\sum\limits_{\bar{a}_{k/k,k-c}}G_v(1,0,\bar{a}_{k/k,k-c},\bar{u}_k)\prod\limits_{s\neq k-c,k}(1-\lambda_{vs}^{a_s}(\bar{u}_s))p(v)dv}{\int_v \sum_{k>W}\sum\limits_{\bar{u}_k}M_v(\bar{u}_k)(1-\lambda_{v,k-c}^{1}(\bar{u}_{k-c}))\sum\limits_{\bar{a}_{k/k,k-c}}G_v(0,1,\bar{a}_{k/k,k-c},\bar{u}_k)\prod\limits_{s\neq k-c,k}(1-\lambda_{vs}^{a_s}(\bar{u}_s))p(v)dv}
\end{equation}
where $M_v(\bar{u}_k)\equiv \lambda_{vk}^{0}(\bar{u}_{k})\prod_{j=1}^k p_v(u_j|\bar{Y}_{j-1}=0,\bar{U}_{j-1}=\bar{u}_{j-1})$ and \sloppy $G_v(a,a',\bar{a}_{k/k,k-c},\bar{u}_k)\equiv p_v(A_k=a|\bar{Y}_{k-1}=0,A_{k-c}=a^{'},\bar{A}_{k/k,k-c}=\bar{a}_{k/k,k-c},\bar{U}_k=\bar{u}_k)\times
\prod_{s=k-c+1}^{k-1}p_v(a_s|\bar{Y}_{s-1}=0,A_{k-c}=a^{'},\bar{A}_{s-1/k-c}=\bar{a}_{s-1/k-c},\bar{U}_{k-c}=\bar{u}_{k-c})\times
p_v(A_{k-c}=a^{'}|\bar{Y}_{k-c-1}=0,\bar{A}_{k-c-1}=\bar{a}_{k-c-1},\bar{U}_{k-c-1}=\bar{u}_{k-c-1})\prod_{s=1}^{k-c-1}p_v(a_s|\bar{Y}_{s-1}=0,\bar{A}_{s-1}=\bar{a}_{s-1},\bar{U}_s=\bar{u}_s)$. If at each time $s$, $A_{s}$ is determined by an independent coin flip with success probability $p$ the bias is approximately $\frac{\int_{v}\sum_{k>W}%
\sum\limits_{\bar{u}_{k}}M_{v}(\bar{u}_{k})\left\{ 1-\lambda
_{v,k-c}^{0}(\bar{u}_{k-c})\right\} p(v)dv}{\int_{v}\sum_{k>W}\sum%
\limits_{\bar{u}_{k}}M_{v}(\bar{u}%
_{k})\left\{ 1-\lambda _{v,k-c}^{1}(\bar{u}_{k-c})\right\} p(v)dv}$. 

%\section*{Appendix 2: Analytic Confirmation of Simulation Results}
%In our first simulation in Section 5.1, the multiplicative bias term from part (i) of Theorem 1 very nearly reduces to:
%\begin{equation}
%\frac{\sum_k\{\lambda^0_k(U_{k-1}=1)p_Up_A(1-\lambda^0_{k-1}(U_{k-1}=1))p_A + \lambda^0_k(U_k=1)p_Up_A\times1\times p_A}{\sum_k\{\lambda^0_k(U_{k-1}=1)p_Up_A(1-\lambda^1_{k-1}(U_{k-1}=1))p_A + \lambda^0_k(U_k=1)p_Up_A\times1\times p_A}
%\end{equation}
%which equals $1.55/1.1\approx 2.8/2$, the bias factor obtained in the simulation. (The only approximation in the above was to ignore the possibility that $U$ is equal to 1 at more than one time in a patient, which is fine since $U$ is rarely 1.)

%Under the fine-binned second simulation data generating process, the bias term for the case-crossover estimator approximately (again, ignoring the possibility that there are multiple hours with $U=1$) reduces to:
%\begin{align*}
%&\frac{\sum_k\sum_{n=1}^{3600}\{\lambda^0(U=1)p_Up_A(1-\lambda^0(U=1))^{3600}p_A(1-\lambda^1(U=1))^{n-1} + \lambda^0(U=1)p_Up_A\times1\times p_A(1-\lambda^1(U=1))^{n-1}\}}{\sum_k\sum_{n=1}^{3600}\{\lambda^0(U=1)p_Up_A(1-\lambda^1(U=1))^{3600}p_A(1-\lambda^0(U=1))^{n-1} + \lambda^0(U=1)p_Up_A\times1\times p_A(1-\lambda^0(U=1))^{n-1}}\\
%&=\frac{[\sum_{n=1}^{3600}(1-\lambda^1(U=1))^{n-1}][(1-\lambda^0(U=1))^{3600}+1]}{[\sum_{n=1}^{3600}(1-\lambda^0(U=1))^{n-1}][(1-\lambda^1(U=1))^{3600}+1]}
%\end{align*}
%which equals $.92\approx 1.84/2$, the bias factor obtained in simulation.
\normalsize

\section*{Appendix 2: Further Details on Sequential
Exchangeability Assumption}

More precisely, equation (4) holds under the assumption (which we assume is true) that the causal DAG in
Figure 1 represents an underlying FFRCIST counterfactual causal model
(Robins, 1986) and thus also under Pearl's NPSEM with independent errors.
See Richardson and Robins (2013) and Shpitser, Richardson, and Robins
(2020).

\section*{Appendix 3: Analysis of Bias Simulations}
\subsection*{Analytic confirmation of the results from the coarse independent exposure simulation}
For $N=100,000$ subjects, we simulated treatments and counterfactual outcomes until the first occurrence
of the outcome according to the following data generating process (DGP):
\begin{equation*}
\begin{split}
& U_{t}\sim Bernoulli(.001);\text{ }\lambda
_{t}^{0}(U_{t-1},U_{t})=min(1/2,.45U_{t-1}+.45U_{t}) \\
& Y_{t}^{0}\sim Bernoulli(\lambda _{t}^{0}(U_{t-1},U_{t}));\text{ }\lambda
_{t}^{1}(U_{t-1},U_{t})=2\lambda _{t}^{0}(U_{t-1},U_{t}) \\
& Y_{t}^{1}\sim Bernoulli(\lambda _{t}^{1}(U_{t-1},U_{t}));\text{ }A_{t}\sim
Bernoulli(.5);\text{ }Y_{t}=A_{t}Y_{t}^{1}+(1-A_{t})Y_{t}^{0}
\end{split}%
\end{equation*}%
The true value of $\beta$ is 2. There are no common causes of treatments and outcomes, treatments are
independent identically distributed and hence exhibit no time trends, and
the outcome is rare when marginalized over $U$. (While the outcome is not rare
when $U_{t}=1$, it is rare that $U_{t}=1$.) Yet the limit of the
case-crossover estimator using the time prior to outcome occurrence as the
control is approximately 2.8.

By Equation (22) in the proof of Theorem 1 in the Appendix (main text, not web), the bias of the case crossover estimator is approximately
\begin{equation}\tag{S1}\label{bias}
\frac{\sum_{k}\int_{v}\sum_{\bar{u}_{k}}\lambda _{vk}^{0}(\bar{u}%
_{k})p_{v}(\bar{U}_{k}=\bar{u}_{k},A_{k-c}=0,A_{k}=1,\bar{Y}_{k-1}=0)p(v)dv}{%
\sum_{k}\int_{v}\sum_{\bar{u}_{k}}\lambda _{vk}^{0}(\bar{u}_{k})p_{v}(\bar{U}%
_{k}=\bar{u}_{k},A_{k-c}=1,A_{k}=0,\bar{Y}_{k-1}=0)p(v)dv}.
\end{equation}
In the DGP above, this expression approximately reduces to:
\small
\begin{equation}\tag{S2}
\frac{\sum_k\{\lambda^0_k(U_{k-1}=1)p_U(1-p_A)(1-\lambda^0_{k-1}(U_{k-1}=1))p_A + \lambda^0_k(U_k=1)p_U(1-p_A)\times1\times p_A}{\sum_k\{\lambda^0_k(U_{k-1}=1)p_Up_A(1-\lambda^1_{k-1}(U_{k-1}=1))(1-p_A) + \lambda^0_k(U_k=1)p_U\times p_A\times1\times(1-p_A)}
\end{equation}
\normalsize
where $p_U$ and $p_A$ denote the Bernoulli parameters of $U_t$ and $A_t$, respectively, in the DGP. The only approximation in the above was to ignore the possibility that $U_t$ is equal to 1 at more than one time $t$ in the same subject, which leads to small approximation error since $U_t$ is rarely 1. Plugging in the parameter values from the DGP, (S2) is equal to $1.55/1.1\approx 2.8/2$, the bias factor obtained in the simulation. 

\subsection*{Analytic confirmation of the results from the fine correlated exposure simulation}
We modified the previous simulation example to add correlations in treatments across time induced by short time bins, and we saw that the bias flips direction. If time bins are interpreted as hours in the previous simulation, they are seconds in this one. Exposure and the unobserved common cause of the outcome are still randomly and independently assigned to one hour intervals as in the previous simulation. This has the effect of inducing (perfect) correlation between treatments in one second time bins within the same hour. The untreated one second discrete hazards are set to preserve the hourly untreated survival probability from the previous simulation, and the multiplicative treatment effect within each one second bin is again set to 2. To formalize, we simulated data according to
\begin{equation*}
\begin{split}
& \tilde{U}_{k}\sim Bernoulli(.001)\text{ for }k\in\{1,\ldots,24\}; U_{kt}=\tilde{U}_{k}\text{ for }k\in\{1,\ldots,24\}, t\in\{1,\ldots,3600\}\\
& \tilde{A}_{k}\sim Bernoulli(.5)\text{ for }k\in\{1,\ldots,24\}; A_{kt}=\tilde{A}_{k}\text{ for }k\in\{1,\ldots,24\}, t\in\{1,\ldots,3600\}\\
& \lambda_{kt}^{0}(\bar{U}_{kt})= 0.000166(U_{kt} + U_{k-1t}); Y_{kt}^{0}\sim Bernoulli(\lambda_{kt}^{0}(\bar{U}_{kt}));\text{ }\lambda_{kt}^{1}(\bar{U}_{kt})=2\lambda_{kt}^{0}(\bar{U}_{kt}) \\
& Y_{kt}^{1}\sim Bernoulli(\lambda_{kt}^{1}(\bar{U}_{kt}));\text{ }Y_{kt}=A_{kt}Y_{kt}^{1}+(1-A_{kt})Y_{kt}^{0}
\end{split}%
\end{equation*}%
where we have indexed `hours' by $k$ and seconds within hours by $t$.

We can again confirm these results analytically. We use the shorthand $\lambda^a(U=1)$ to denote the hazard at time $kt$ if $U_{k-1t}=1$ or if $U_{kt}=1$, ignoring for the sake of convenient approximation the possibility that there are multiple hours with $\tilde{U}_k=1$. The bias term for the case-crossover estimator (S1) approximately (again, under the simplifying assumption that there are not multiple hours with $\tilde{U}_k=1$) reduces to:
\tiny
\begin{align*}\label{fine_bias}
&\frac{\sum_{k=2}^{24}\sum_{t=1}^{3600}\{\lambda^0(U=1)p_U(1-p_A)(1-\lambda^0(U=1))^{3600}p_A(1-\lambda^1(U=1))^{n-1} + \lambda^0(U=1)(1-p_A)\times1\times p_Up_A(1-\lambda^1(U=1))^{t-1}\}}{\sum_{k=1}^{24}\sum_{t=1}^{3600}\{\lambda^0(U=1)p_Up_A(1-\lambda^1(U=1))^{3600}(1-p_A)(1-\lambda^0(U=1))^{t-1} + \lambda^0(U=1)p_A\times1\times p_U(1-p_A)(1-\lambda^0(U=1))^{t-1}}\\
&=\frac{[\sum_{t=1}^{3600}(1-\lambda^1(U=1))^{t-1}][(1-\lambda^0(U=1))^{3600}+1]}{[\sum_{t=1}^{3600}(1-\lambda^0(U=1))^{t-1}][(1-\lambda^1(U=1))^{3600}+1]}.
\end{align*}
\normalsize
where $p_U$ and $p_A$ denote the Bernoulli parameters of $\tilde{U}_k$ and $\tilde{A}_k$, respectively, in the DGP. Plugging in the parameter values from the DGP, this expression is equal to $.92=1.84/2$, the bias factor obtained in simulation. 

Examining this bias approximation, we can see how the bias gets pushed toward the null. Selection on surviving the control hour when $\tilde{U}_{k-1}=1$ leads to $1-\lambda^0(U=1)$ terms in the numerator and $1-\lambda^1(U=1)$ terms in the denominator. We argued in Section 5.1 that the discrepancy between these terms pushes the bias away from the null, as in the first simulation. Selection on surviving the portion of the case hour preceding the occurrence of the event leads to $1-\lambda^1(U=1)$ terms in the numerator and $1-\lambda^0(U=1)$ terms in the denominator, which by analogous reasoning pushes the bias toward the null. Selection on surviving the control hour only enters into the formula if $\tilde{U}_{k-1}=1$, since risk is 0 whenever $U$ is 0. Selection on surviving the case hour preceding the event, however, occurs whether $\tilde{U}_k=1$ or $\tilde{U}_{k-1}=1$. This explains how terms pushing the bias factor toward the null outweigh terms pushing the bias factor away from the null in this example.

\section*{Appendix 4: Possible Violations of Assumptions in the MI Study}
It might be illustrative to assess our simplified version of Mittelman et al.'s (1993) seminal study on the impact of exercise on MI through the lens of our analysis. Recall that in the simplified version data collection occurs over the course of a single Sunday, so we take the baseline for the underlying cohort population of interest to be midnight of the preceding Saturday morning. It is worth noting that this study highlights the benefit of the case-only nature of the design in that it would be very difficult to collect the data required to perform a cohort study targeting an equivalent question. 

\textit{No post-baseline confounding (Sequential Exchangeability, assumption (4)).} We discussed the example of drinking coffee as a possible violation of this assumption. Caffeine might increase energy and encourage exercise and also independently increase the risk of MI. We might reasonably hope that confounders of this sort (short term encouragements to exercise that are also associated with MI) are weak.   

\textit{No direct effect of treatment on later outcomes (UV-Transient Hazards, assumption (5)).} It is possible that exercise has a cumulative effect on the outcome. Two consecutive hours of exercise might cause an MI in some subjects for whom just one hour would not. Perhaps extended vigorous exercise is rare enough that the cumulative effect of exercise does not seriously impact results or their interpretation. Delayed effects of exercise are not thought to be significant.

\textit{No time modified confounding (assumption (8)).} It might be that within levels of certain baseline confounders, exercise is more probable on Saturday (or Sunday) afternoon and MIs are more (or less) probable than average, even if marginal probability of exercise in the full cohort is equal on the two days. If the control hour is taken to be 24 hours before the MI, this scenario would induce bias. As a strained, stylized, and purely illustrative example, in the United States men are more likely to have MIs than women and also more likely to be fans of the National Football League (NFL). Suppose our study takes place the weekend of Super Bowl Sunday, which is the day of the NFL championship game. Then men in our study would be particularly less likely to exercise on Sunday than Saturday, which could lead to excessive $unexposed$ MIs on Sunday, making exercise appear a less potent cause of MI than it is. %perhaps the lifestyle of a stereotypical NFL fan (beer, cholesterol) makes MIs more likely and also makes exercise during the afternoon on Sunday (when most NFL games are played) less likely than Saturday. This dynamic could lead to excessive $unexposed$ MIs on Sunday, making exercise appear a less potent cause of MI than it is.

\textit{No time trends in treatment (assumption (9)).} Marginal probability of recent exercise varies greatly by time of day. If the control time is chosen appropriately (e.g. exactly 24 hours before the MI), then approximate pairwise exchangeability may hold. But perhaps there are reasons why exercise is generally more or less common on Sunday than Saturday (e.g. church or football games).

Under the above assumptions (in addition to Consistency), the case-crossover could reasonably be applied to test the causal null hypothesis. To interpret the case-crossover point estimate, additional considerations are required.

\textit{Rare outcome (assumption (7)).} The outcome must be rare within all levels of the baseline confounders, exposure, and common causes of the outcome. MIs are certainly rare marginally at the level of a day, and probably also rare across levels of baseline confounders and exposures. However, we mentioned that perhaps causes of the outcome such as presence of a clot could make the outcome common, particularly under exposure. This would induce bias of the sort seen in the simulation in Section 5.1. 

\textit{Constant causal hazard ratio (assumption (6)).} It is highly unlikely that the multiplicative effect of exercise across hour and covariate levels is constant. While true under the null, this is a very strong assumption if the null does not hold. We saw in Section 5.2 that it is difficult to interpret the point estimate if the effect is heterogeneous. 

\end{document}